\documentclass[amssymb, nobibnotes, aps, prd]{revtex4-1}
\usepackage[utf8]{inputenc}
\usepackage{epsfig}
\usepackage{amsmath}
\usepackage{amsfonts}
\usepackage{amssymb}
\usepackage{graphicx}
\usepackage{colordvi}
\usepackage{amsthm}
\usepackage{slashed}
\usepackage{bbm}
\usepackage{color}
\usepackage{amsfonts}
\newtheorem{Def}{Definition}
\newtheorem{Prop}{Proposition}
\newtheorem{Thm}{Theorem}

\usepackage{array}

\begin{document}
\begin{flushleft}
KCL-PH-TH/2016-59
\end{flushleft}

\title{Noncommutative geometrical origin of the energy-momentum dispersion relation}

\author{
  A.\ Watcharangkool\footnote{email address:
 apimook.watcharangkool@kcl.ac.uk}
and
M.\ Sakellariadou\footnote{email address:
    mairi.sakellariadou@kcl.ac.uk}}

 \affiliation{Department of Physics, King's College London, University
  of London, Strand WC2R 2LS, London, United Kingdom}

\begin{abstract}
  We investigate a link between the energy-momentum dispersion
  relation and the spectral distance in the context of a
  Lorentzian almost-commutative spectral geometry, defined by
  the product of Minkowski spacetime and an internal discrete
  noncommutative space. Using the causal structure, the
  almost-commutative manifold can be identified with a pair of
  four-dimensional Minkowski spacetimes embedded in a five-dimensional
  Minkowski geometry. Considering fermions travelling within the
  light cone of the ambient five-dimensional spacetime, we then
    derive the energy-momentum dispersion relation.

\end{abstract}
%\date{\today}
\maketitle

\section{Introduction}
The framework of noncommutative geometry (NCG) offers a generalisation
to the notion Riemannian geometry, replacing manifolds with algebras of
bounded operators on Hilbert spaces~\cite{ncg-book2}. The
  formalism was first used for commutative $C^*$-algebras, while then
  was extended to {\sl spaces} characterised by a
  noncommutative algebra of coordinates. Extending all basic
geometric notions from ordinary manifolds to noncommutative spaces is
a fundamental aspect of noncommutative geometry. In such a
framework, all information about a physical system is encoded within
the algebra of operators in a Hilbert space, with the action expressed
in terms of a generalised Dirac operator. Following this approach, all
fundamental forces in physics can be considered on an equal
footing, namely as curvature on a noncommutative manifold, leading to
a purely geometric explanation for the Standard Model of
particle physics~\cite{Chamseddine:2006ep}. In addition, this
  approach implies an equivalent formulation for the distance on a
manifold, defined as a set of pure states of a commutative
$C^*$-algebra. For example, on a manifold where points are identical
to pure states of commutative $C^*$-algebra, the geodesic distance between
points on the manifold is completely determined by spectral data of a
Dirac operator
\begin{equation}
  d(x,y)={\rm sup}\{|\omega_x(f)-\omega_y(f)|: f\in A, ||[-i\slashed{\nabla},f]||\leq 1\}~, \label{eq:dis-form}
\end{equation}
where $A$ is a commutative pre-$C^*$-algebra, $\omega_{x,y}$ are pure
states of the algebra defined by $\omega_x(f):=f(x)$, and $-i\slashed{\nabla}$ is the
Dirac operator associated with the spin connection, playing the role of the inverse of the line element
$ds$ (where $ds=\sqrt{g_{\mu\nu}dx^\mu dx^\nu}$).
Equation~(\ref{eq:dis-form}) above is known as \textit{spectral
  distance formula} or \textit{Connes' distance formula}. As a
distance function between pure states, the above expression makes
perfect sense when one generalises the commutative algebra to a
noncommutative one, however the physical meaning of this quantity is
not clear in the noncommutative regime. It has been
shown~\cite{ncgdis} that in an almost-commutative manifold, the
spectral distance resembles the geodesic distance in a higher
dimension manifold, but extracting physical meaning of this result is
nontrivial.

An important issue of NCG is the lack of its Lorentzian version, which
is the geometry of our physical spacetime. Strictly speaking, there is
no particle physics model from NCG, but a model inspired by NCG. To
investigate the energy-momentum dispersion relation, which is obtained
in the framework of a relativistic theory, one may have to include the
notion of causal structure into the geometry. Thus, in what follows,
we will incorporate generic features about Lorentzian
noncommutative geometry~ \cite{NCG-L,NCG-cone,Koen, KRennie, Alex}.

The rest of this paper is organised as follows: In the Section~{\rm
  II}, we discuss some general properties of the spectral triple and
the spectral distance formula. In Section~{\rm III}, we state
the definition of Lorentzian spectral triple, which will be used
throughout this paper, and elaborate on the notion of causal
structure. In Section~{\rm IV}, we investigate the link between the
distance formula and the energy-momentum dispersion relation. We
conclude  in Section~{\rm V}.
%---------------------------------------------------------------
\section{almost-commutative geometry and distance formula}
\subsection{Spectral Triples}
The spectral triple is a collection of data $(A,\mathcal{H},D)$, where
$A$ is a dense subalgebra of a $C^*$-algebra (pre $C^*$-algebra) acting as a subalgebra of bounded operators on
a Hilbert space $\mathcal{H}$, and $D$ is a Dirac operator (densely
defined self-adjoint operator with compact resolvent). It can be seen
as a generalised notion of geometry: if $A$ is a unital commutative
algebra, namely if we have a commutative spectral triple, then one can
reconstruct the compact Riemannian spin manifold $M$, such that
$A\simeq C^\infty(M)$~\cite{C-Man}. It is this duality between a
commutative $C^*$-algebra and the algebra of smooth functions on a Riemannian
manifold that inspired the notion of noncommutative
geometry: given a
noncommutative algebra $A$, one may think of a noncommutative geometry
as a space $X$ for which $A$ is the coordinate algebra.

In addition, one considers a real structure $J$ and a grading operator
$\gamma$ (we refer the reader to Ref.~\cite{NCG-book} for details),
which are crucial for the construction of spin manifold and obtain the
Standard Model of high energy physics from noncommutative spectral
geometry.

Let $M\times F$, where $M$ is a four-dimensional Riemannian spin
manifold and $F$ an internal noncommutative space, define an
almost-commutative manifold. Its spectral triple $(A,\mathcal{H},D)$
is given by the algebra
\begin{equation}
C^\infty(M)\otimes A_F:=C^\infty(M)\otimes\Big(\bigoplus^n_{k=1}A_k\Big)~,
\end{equation}
with finite-dimensional algebra (not necessarily commutative)  $A_F$,
 Hilbert space $L^2(M,S)\otimes\mathcal{H}_F$ and  Dirac
operator $-i\slashed{\nabla}\otimes{\rm Id}_F+\gamma^5\otimes D_F$,
where $H_F$ is a finite-dimensional Hilbert space and $D_F$  a
self-adjoint matrix (Dirac operator).

Choosing appropriately the algebra of the internal space $F$ as
\begin{equation}
A_F=\mathbbm{C}\oplus\mathbbm{H}\oplus M_4({\mathbbm{C}})~,
\end{equation}
and applying the spectral action, which is basically the trace of the
heat kernel of the Dirac operator, one obtains an effective
description of the Standard Model~\cite{Walter}.

%%%%%%%%%%%%%%%%%%%%%%%%%%%

\subsection{Inner fluctuations}
The symmetry in an almost commutative manifold is the automorphism
group of the algebra
\begin{equation}
{\rm Diff}(M\times F):={\rm Aut}(C^\infty(M,A_F))~,
\end{equation}
since the diffeomorphism group, which is the symmetry group on a
manifold, is isomorphic to the automorphism of the algebra of smooth functions,
${\rm Diff}(M)\simeq {\rm Aut}(C^\infty(M))$. Being interested in the
automorphism that would lead to the symmetries of the Standard Model,
let us consider the inner automorphism $\alpha_u$,
 characterised by a unitary element of the algebra
\begin{equation}
\alpha_u(a)\mapsto uau^*~,
\end{equation}
where $u \in \mathcal{U}(A)$. Since the unitary equivalence is an
important element for the physics of the Standard Model, we need to
incorporate it in the spectral action. To do so, we define an algebra
$B:=\alpha_u(A)\simeq A$ as a unitary equivalent algebra, and find its
corresponding spectral triple $(B,\mathcal{H}',D')$, which involves
the notion of Morita equivalence. The Morita equivalence between two
$C^*$-algebras $B$ and $A$ implies the existence of a projective right
$C^*$-module $\mathcal{E}$ (we refer the reader for more details on
$C^*$-module in Ref.~\cite{NCG-book}) such that
\begin{equation}
B={\rm End}_A(\mathcal{E})~.
\end{equation}
Note that, in the case where the algebra has both left- and right-action on the Hilbert space, the definition of
Morita equivalence requires a bimodule.

Since that algebra is the ${\rm End}_A(\mathcal{E})$, the natural
choice for the Hilbert space of the new triple is
$\mathcal{H}':=\mathcal{E}\otimes_A\mathcal{H}$, it remains to
  choose the  Dirac operator.  Suppose there exists a Hermitian
connection $\nabla:\mathcal{E} \rightarrow
\mathcal{E}\otimes\Omega^1_D$ satisfying the conditions
\begin{align}
\nabla(\xi a)=&(\nabla\xi)a+\xi\otimes da~, ~\forall\xi\in \mathcal{E}, 
a\in A~, \\
d\langle\xi,\eta\rangle_A=& \langle\xi,\nabla\eta\rangle_A-\langle\nabla\xi,
\eta\rangle_A~, \forall\xi,\eta \in \mathcal{E}~,
\end{align}
where $da:=[D,a]$,  $\Omega^1_D$ is the algebra of one-forms and $\langle \cdot , \cdot\rangle_A:
\mathcal{E}\times\mathcal{E}\rightarrow A$ denotes the Hermitian
product. Then the Dirac operator can be defined by
\begin{equation}
D'(\xi\otimes\eta)=\xi \otimes D\eta+(\nabla\xi)\eta.  
\end{equation}
For $B:=\alpha_u(A)\simeq A$ , we have
$\mathcal{E}=A$, hence the Dirac operator is
 \begin{align}
 D'(1_A\otimes\eta)=1_A\otimes D\eta+(d1_A)\eta~.
 \end{align} 
 When $d1_A=[D,1_A]\not=0$ the Dirac operator $D'$ is
 $D'=D+\mathcal{B}$, where $\mathcal{B}$ is a self-adjoint element of
 $\Omega^1_D(A)$ and plays the role of gauge potential.  Given the
 charge conjugation operator, the Dirac operator reads
\begin{align}
D'=D+\mathcal{B} +\epsilon'J\mathcal{B}J^{-1}~,
 \end{align} 
 called the \textbf{inner fluctuation}, with $J$ a real structure (an
 antilinear isometry $J:{\cal H}\rightarrow {\cal H}$) and the number
 $\epsilon'\in\{-1,1\}$ a function of $n$ mod 8.

%%%%%%%%%%%%%%%%%%%%%
\subsection{Spectral Distance Formula}
We have previously seen the spectral distance formula in the case of a
commutative spectral triple, where elements of the algebra are just
smooth functions. Since the formula is defined purely from spectral
data, it is still valid for a noncommutative spectral triple. Hence,
\begin{equation}
d(\omega,\omega')=\sup\{|\omega(a)-\omega'(a)|~:~a\in A,~\|[D,a]\| \leq 1\}~, \label{pure-dis}
\end{equation}
where $\omega,\omega'\in \mathcal{P}(A)$ are pure states of the
algebra $A$, having in mind a generalised notion of points. Note that, although
the distance formula exists, the notion of distance between any two pure states
is well-defined only when $d(\omega,\omega')< \infty$. Even though
 we consider a spectral triple in which the formula \eqref{pure-dis} gives finite distance, 
 the meaning of the distance between pure states in an abstract noncommutative
space is still quite difficult to understand. Nevertheless, in the case of an almost-commutative manifold, its
pure states are isomorphic to the points on the product space,
i.e. $\mathcal{P}(A)\cong M\times F$~\cite{Walter}. In the case that $F$ is a finite space, the geodesic distance squared
between $(x, e_i) $ and $(y, e_j) $, for $e_i, e_j \in F$ is given
by~\cite{ncgdis}
\begin{equation}
d^2(x\times e_i, y\times e_j)=d_M^2(x,y)+d_F^2(e_i,e_j)~, \label{DPyth}
\end{equation}
where $d_M(x,y)$ is the geodesic distance on $M$ and $d_F(e_i,e_j)$ stands
for the shortest distance between internal states $ e_i $ and $ e_j$.
This Pythagorean theorem allows one to embed the almost-commutative
manifold $M\times F$ in a $(n+1)$-dimensional Riemannian manifold
$M\times \mathbbm{R}$. The metric of the almost-commutative manifold
inherited from the ambient  $(n+1)$-dimensional manifold is
\begin{equation}
g_{ab}=\left( \begin{array}{cc} g_{\mu\nu} & 0 \\ 0 & 1/d^2(e_i,e_j) 
\end{array}\right)~,\label{eq:5D-metric}
\end{equation}
where $a, b \in \{0,1, 2, 3, 4\}$ (namely they refer to the
almost-commutative manifold), and Greek indices $\mu,\nu \in \{0, 1, 2,
3\}$.  The physical meaning of the Dirac operator, as discussed
earlier, implies 
\begin{equation}
ds^{-2}\Big|_{M\times F}=D^2=-\slashed{\nabla}^2+D^2_F~,
\end{equation}
and hence $D$ satisfies the Pythagorean theorem.

For the simple model of a two-sheet space $M\times\{0,1\}$ with
discrete spectral triple $(A_F, \mathcal{H}_F,D_F)$, given by
\begin{equation}
  A_F=\mathbbm{C}\oplus\mathbbm{C}\ ,~\mathcal{H}_F=\mathbbm{C}^2\ ,~ 
D_F=\left(\begin{array}{cc} 0 & m \\ m^* & 0 \end{array}\right)~,
 \label{eq:two-sheeted}
\end{equation}
where $m\in \mathbbm{C}$ is a non-zero complex parameter, we have
$d_F(0,1)=1/|m|$. So in this case $D^2_F=|m|^{2}\mathbbm{1}_2$. Note
that, although $|m|$ is a constant in the two-sheet space, it can be
a function of $x\in M$ if one considers an almost-commutative space with
inner fluctuations.

In what follows, we restrict our study to the two-sheet space, since
it was shown in Ref.~\cite{ncgdis} that if the internal space of almost-commutative manifold is discrete,
 then one can reduce the distance formula in an almost-commutative manifold into that of
a two-sheet geometry.

%---------------------------------------------------------------
\section{Lorentzian spectral triple}

Although noncommutative geometry has been applied to a relativistic
theory like the Standard Model, the definition of a Lorentzian
spectral triple remains an open question, the reason mainly being the
lack of manifold reconstruction theorem analogous to Connes'
reconstruction theorem for a commutative spectral
triple~\cite{C-Man}. Nevertheless, there are a few similar definitions 
of Lorentzian spectral triples in the literature \cite{NCG-cone, Koen, Alex, KRennie}.
  In this paper we adopt the definition proposed by \cite{NCG-cone},
 which will be sufficient to define a causal structure. Moreover, for a commutative case that is
constructed from a globally hyperbolic manifold, one can define a distance 
formula (which will be defined in the next section) similar to the spectral distance formula. 
The Lorentzian version of spectral distance formula was proposed in \cite{Full-pf},
it was proved that the formula leads to the geodesic distance in Minkowski space.
\begin{Def} \textbf{Lorentzian spectral triple}\\
  A Lorentzian spectral triple is given by
  $(A,\tilde{A},\mathcal{H},D,\mathcal{J})$, where
\begin{itemize}
\item $A$ is a non-unital dense $*$-subalgebra of a $C^*$-algebra, and
  $\tilde{A}$ its preferred unitalisation
\item $\mathcal{H}$ is a Krein space with an indefinite product
  $(\cdot,\cdot)$
\item $\mathcal{J}$ is a bounded self-adjoint symmetry operator,
  $\mathcal{J}=\mathcal{J}^*,~\mathcal{J}^2=1$, commuting with $A$.
  The role of $\mathcal{J}$ -- dubbed as {\sl fundamental symmetry} or
  {\sl signature operator} -- is to turn the Krein space
  $\mathcal{H}$ into a Hilbert space. Note that, $\mathcal{H}_\mathcal{J}$ is the same space as
  $\mathcal{H}$ with positive definite inner product $\langle\cdot,
  \cdot\rangle:=(\cdot,\mathcal{J}\cdot)$, hence a Hilbert space.
 \item $D$ is a densely defined operator on $\mathcal{H}_\mathcal{J}$ such that
 \begin{itemize}
 \item $D=-\mathcal{J}D^*\mathcal{J}=:-D^+$ i.e. it is Krein
   anti-self-adjoint on $\mathcal{H}$
 \item $\forall a \in \tilde{A},~ [D,a]$ extends to a bounded operator
   on $\mathcal{H}_\mathcal{J}$
 \item $\forall a \in A,~ a(1+\langle D \rangle)^{-1/2}$ is compact on
   $\mathcal{H}_{\mathcal{J}}$, where $\langle D\rangle^2 :=
   \frac{1}{2}(DD^*+D^*D)$
 \end{itemize} 
\item there exists a densely defined self-adjoint operator
  $\mathcal{T}$ with ${\rm Dom}D \cap {\rm Dom}\mathcal{T}$ dense in
  $\mathcal{H}_\mathcal{J}$ such that
 \begin{itemize}
 \item $(1+\mathcal{T}^2)^{-1/2} \in \tilde{A}$
 \item $\mathcal{J}=-N[D,\mathcal{T}]$ for some positive element $N\in
   \tilde{A}$.
 \end{itemize}
\end{itemize} 
\end{Def}
Let us consider the Lorentzian spectral triple~\cite{NCG-L}
\begin{equation}
(C^\infty_0(M),C^\infty_b(M), L^2(M,S), -i\slashed{\nabla})~,
\end{equation}
where $M$ is a globally hyperbolic Lorentzian manifold with signature
$(-,+,+,+)$ , $C^\infty_0(M)$ is the algebra of smooth functions
vanishing at infinity, and $C^\infty_b(M)$ is for the space of smooth
bounded functions on the manifold. The Krein $L^2(M,S)$ is the space of 
square integrable smooth sections of the spinor bundle. 
The Dirac operator is defined by $-i\slashed{\nabla}:=-i\gamma^\mu\nabla_\mu$, 
where $\nabla_\mu$ is the spin connection on $M$. Note that we choose the representation
of the gamma matrices such that
\begin{equation}
(\gamma^0)^*=-\gamma^0 ,~~~ (\gamma^k)^*=\gamma^k~,
\end{equation}
where $k=1,2,3$, and satisfy the relation
\begin{equation}
\{\gamma^\mu,\gamma^\nu\}=2g^{\mu\nu}\mathbbm{1}_4~.
\end{equation}

The fundamental symmetry $\mathcal{J}$ can be derived from the lapse
function $N$ and the global time function $\mathcal{T}$, as follows:
For a globally hyperbolic Lorentzian manifold $M$, there exists a
global smooth time function $\mathcal{T}$ on $M$ such that the line
element of the manifold $M$ splits as
\begin{equation}
 ds^2=-Nd\mathcal{T}^2+ds^2_\mathcal{T}~,
 \end{equation}
 where $ds^2_\mathcal{T}$ is the line element on the Cauchy
 hypersurface $\Sigma_\mathcal{T}$ at constant time $\mathcal{T}$ and
 $N$ is the lapse function.  The fundamental symmetry in terms of $N$
 and $\mathcal{T}$ is $\mathcal{J}=-N[D, \mathcal{T}]=-iN\gamma^0$; a condition that
 guarantees the Lorentzian signature.
 
 To include a causal structure into the algebra, one defines a set of
 real-valued functions which are non-decreasing along a
 future-directed causal curve:
\begin{equation}
  \mathcal{C}=\{f \in C^\infty_b(M) : ~f(x) \leq f(y) ~{\rm iff}~ x\preceq y,~ 
\forall x,y \in M\}~.
\end{equation}
The set $\mathcal{C}$ is called the \textbf{causal cone} and its
elements are \textbf{smooth bounded causal functions}. In a globally hyperbolic
spacetime $(M,g)$, the geodesic distance coincides with the Lorentzian
distance function~\cite{Lorentz-dis}
\begin{equation}
  d(x,y)=\inf\Big\{ f(y) - f(x) \Big|~ f \in \mathcal{C}~,~{\rm ess ~sup}~ 
g(\nabla f,\nabla f) \leq -1~~, ~~ \forall x,y \in M\ \mbox{with} 
~x\preceq y\Big\}~.\label{eq:sp-dis}
\end{equation}
In the following, we highlight the definition of the causal cone
expressed in terms of the spectral triple~\cite{Full-pf, NCG-L}.

\begin{Prop} Let $(A,\tilde{A},\mathcal{H},D,\mathcal{J})$ be a
  commutative Lorentzian spectral triple constructed from a globally
  hyperbolic manifold. Then $f\in \tilde{A}$ is a causal function iff
\begin{equation}
(\psi, [D,f]\psi) \leq 0~,~\forall \psi \in \mathcal{H}~.
\end{equation}
\end{Prop}
\noindent
This can be generalised to a noncommutative spectral triple by
replacing $A$ with a noncommutative algebra~\cite{NCG-L}.

For simplicity, let us consider a Minkowski spacetime, denoted by
$\mathcal{M}$, as the globally hyperbolic spacetime. In a
four-dimensional Minkowski spacetime, any two points $x,y \in
\mathcal{M}$ can be connected by a spacelike curve, i.e. a curve
$\gamma:[0,1]\rightarrow \mathcal{M}$ such that
$g(\dot{\gamma},\dot{\gamma}) > 0$ along the curve. However, some of
these points can also be connected by a causal curve,
i.e. $g(\dot{\gamma},\dot{\gamma})\leq 0$ everywhere along the curve;
these points are called causally related and are denoted by $x
\preceq$ y.

Consider two points $x, y$ in the Minkowski four-dimensional
spacetime $\mathcal{M}$, with signature $(-, +, +, +)$, connecting
through a curve $\gamma$. We define the \textbf{extremal length
  squared} as
\begin{align}
  L^2(x,y):=\left\{ \begin{array}{ll} -{\rm
        sup}\{~l(\gamma)^2:=\left(\int_{\gamma}{\sqrt{-g(\dot{\gamma},\dot{\gamma})}}d\tau\right)^2
      ~|~ g(\dot{\gamma},\dot{\gamma})\leq 0\}~&,  ~ x \preceq y \\ \\
      {\rm
        sup}\{~l(\gamma)^2:=\left(\int_{\gamma}{\sqrt{g(\dot{\gamma},\dot{\gamma})}}d\tau\right)^2~|~g(\dot{\gamma},\dot{\gamma})
      > 0\}~&, ~ x \npreceq y ~.\end{array}\right. \label{eq:EL}
\end{align}
Since Minkowski spacetime is flat, $L^2(x,y)=-(x_0-y_0)^2+\|\mathbf{x}
- \mathbf{y} \|^2$, which is zero or negative for two causally related
points and strictly positive otherwise. Notice that, using $L^2(x,y)$
above, we can differentiate between points which are connected by a null curve 
and those which are not causally related. However, the distance
defined by
\begin{align}
  d(x,y)=\left\{ \begin{array}{ll} \sqrt{-L^2(x,y)} &, ~ x \preceq y
      \\ \\ 0~&, ~ x \npreceq y \end{array}\right.~
\end{align}
vanishes for both space-like and light-like separation.  

%--------------------------------------------------------------------------
\section{Energy-Momentum dispersion relation 
almost commutative
  spectral geometry}
In the previous section we have seen that the commutative Lorentzian
spectral triple
$(C^\infty_0(\mathcal{M}),C^\infty_b(\mathcal{M}),L^2(S,\mathcal{M}),
\slashed{\partial})$,
yields a spectral distance equivalent to the geodesic distance for
Minkowski spacetime. Next, we shall define a distance function for an
almost commutative geometry, namely the product of this Lorentzian
spectral triple with a finite spectral triple, and  examine the
implications of the proposed distance function definition for
relativistic particles.

%%%%%%%%%%%%%%%%%%%%%%%%%%%%%%
\subsection{Causal structure and distance}
Consider a two-sheet space, defined by the tensor product of a
commutative Lorentzian spectral triple and a discrete spectral triple
$(A_F,\mathcal{H}_F,D_F)$, as in Eq.~(\ref{eq:two-sheeted}). Following
Ref.~\cite{NCG-cone}, one can  define a causal structure on the
space of states $\mathcal{S}(\tilde{A})$ of the two-sheet space,
using only the spectral data of the almost commutative manifold; we
highlight the procedure below.

\begin{Def} Let $\mathcal{C}=\{a\in \tilde{A}~ |~ a=a^*, ~(\psi, [D,
  a]\psi) \leq 0, \forall \psi \in \mathcal{H}\}$ such that~ ${\rm span}_\mathbbm{C}(\mathcal{C})=\tilde{A}$. Two states $\omega,
  \omega' \in \mathcal{S}(\tilde{A})$ are causally related
  i.e. $\omega \preceq \omega'$ iff for any $a \in \mathcal{C}$, one
  has
\begin{equation}
\omega(a)\leq \omega'(a).
\end{equation}
\end{Def}
\noindent
Let us denote by $\mathcal{P}(A)$ the set of pure states of the
algebra $A$, defined as the union of
$\mathcal{M}_0:=\mathcal{M}\times\{0\}$ and
$\mathcal{M}_1:=\mathcal{M}\times\{1\}$, hence the name of two-sheet
spacetime.  Thus, one may think of having two sheets of
four-dimensional Minkowski spacetimes embedded in a five-dimensional
one. Since we are interested in the causal relation between points on
$\mathcal{M}_0$ and $\mathcal{M}_1$, we consider a particular type of
mixed states $\omega_{x,\xi}\in
\mathcal{N}(A):=\mathcal{M}\times[0,1]\subset \mathcal{S}(A)$ defined
by
\begin{equation}
\omega_{x,\xi}(a\oplus b)=\xi a(x)+(1-\xi)b(x),
\end{equation}
for $a,b \in C^\infty_0(\mathcal{M})$. Such states $\omega_{x,\xi}$
can be considered as covering the area between the two sheets. The
pure states in ${\cal M}(A)$ can be recovered with the coice
$\xi=0$ or $\xi=1$.

\begin{Thm} The two states $\omega_{x,\xi},~ \omega_{y,\eta} \in
  \mathcal{N}(A)$ are causally related if and only if $x \preceq y$ on
  $\mathcal{M}$ and
\begin{equation}
l(\gamma) \geq \frac{|{\rm arcsin}\sqrt{\eta}-{\rm arcsin}\sqrt{\xi}|}{|m|}~,
\end{equation}
\end{Thm}
\noindent
where $l(\gamma)$ represents the length of a causal curve $\gamma$
going from $x$ to $y$ on the manifold $\mathcal{M}$.

The above theorem~\cite{NCG-cone} implies that if the discrete Dirac
operator is trivial, i.e. $m=0$, the causal relation holds only when
$\xi=\eta$. Hence, the extremal length squared between two points
$(x,0), (y,0) \in \mathcal{M}_0$ is
\begin{equation}
  L^2(x,y)=-\sup_{\gamma}l^2(\gamma)=
-(x_0-y_0)^2+\|\mathbf{x} - \mathbf{y} \|^2~,
\end{equation}
where $\gamma$ denotes a causal curve.

If $m\not= 0$, any two points $(x,0)\in \mathcal{M}_0$ and $(y,1)\in
\mathcal{M}_1$ are causally related iff there is a causal curve
$\gamma$ connecting $x$ and $y$ such that
\begin{equation}
l(\gamma)\geq \frac{\pi}{2|m|}~, \label{eq:joint}
\end{equation}
implying
\begin{align}
-\sup_{\gamma}l^2(\gamma)+\frac{\pi^2}{4|m|^2} \leq 0~.
\end{align}
For any $(x,i), (y,j)\in \mathcal{M}\times \{0,1\}$ with $i,j \in
\{0,1\}$ we define
\begin{align}
\label{def}
L^2_m[(x,i),(y,j)]=\left\{ \begin{array}{ll} \frac{4}{\pi^2}L^2(x,y)+\frac{1}{|m|^2}&,
 ~i\not= j \\ \\
                                                           \frac{4}{\pi^2}L^2(x,y) &, ~i=j 
\end{array}\right.
\end{align}
One notices that Eq.~\eqref{def} is the Lorentzian version of the
Pythagorean theorem Eq.~\eqref{DPyth}.

From Eq.~(\ref{eq:EL}), we see that the above defined function, which
we also call {\bf extremal length squared} on $\mathcal{M}\times
\{0,1\}$, is negative semi-definite when the points $(x,i)$ and $(y,
j)$ are causally related, and positive otherwise.  Combining the
definition (\ref{def})  and Theorem 1, one obtains a criterion
for any two points (pure states) to be causally related.
\begin{Prop}
  The pure states $(x,i)$ and $(y,j)$, defined on an almost-commutative manifold, are said
  to be causally related if and only if $x \preceq y$ on ${\cal M}$ and
\begin{equation}
L^2_m[(x,i),(y,j)]\leq 0~.
\label{causal-str}
\end{equation}
\end{Prop}
\noindent
We will refer to the above condition as the {\bf causal structure}.

One notices that the causal structure of the two-sheet space is
exactly the same as the one of a pair of four-dimensional Minkowski
spacetimes embedded in a five-dimensional one
$(\mathcal{M}_5:=\mathcal{M}\times [0,1])$,
with $1/|m|$ denoting the separation between the two four-dimensional
manifolds. The metric of the five-dimensional Minkowski spacetime
$\mathcal{M}_5$ reads
\begin{equation}
  g_{ab}=\left(\begin{array}{cc}  \eta_{\mu\nu} & 0 \\ 0 &
 1/|m|^2 \end{array}\right)~, \label{eq:5-met}
 \end{equation}
 where $\mu,\nu$ are the spacetime indices in Minkowski spacetime,
 which being flat is denoted by $\eta_{\mu\nu}$. The metric
 (\ref{eq:5-met}) can be seen as a wick-rotated version of
 (\ref{eq:5D-metric}).

 Using metric (\ref{eq:5-met}), any two points in the two-sheet
 spacetime are causally related provided they are causally related in
 $(\mathcal{M}_5,g)$. The line element in $\mathcal{M}_5$ is
\begin{align}
ds^2=&~g_{ab}dx^a dx^b = \eta_{\mu\nu}dx^\mu dx^\nu + 
\frac{1}{|m|^2}dx^2_F \nonumber \\
       =&~ds^2_\mathcal{M}+ds^2_F~,
\end{align}
where $dx_F$ is the infinitesimal of the interval $[0,1]$. 

 Making the appropriate choice for the Dirac operator ${\cal D}$ in
 $M_5$, such that
\begin{equation}
{\cal D}^2=-\slashed{\nabla}^2-|m|^2\frac{\partial^2}
{\partial x^2_F}~,
\label{calD}
\end{equation}
the spectral distance expression (\ref{eq:sp-dis}) for a globally
hyperbolic manifold, implies the geodesic expression as the one
derived from the metric (\ref{eq:5-met}).  To specify our notation,
let us remark that ${\cal D}$ is defined by Eq.~(\ref{calD}), whereas
$D$ will refer to the Dirac operator as defined for an almost
commutative manifold.

The Lorentzian version of the spectral distance formula is still
applicable on the two-sheet space, since it is a submanifold of
  $\mathcal{M}_5$. Note that, to recover the $D^2$ operator as
defined for an almost-commutative Lorentzian manifold, one chooses the
boundary condition for a spinor in a five-dimensional Minkowski space
such that for any $\phi \in L^2(M_5, S)$
\begin{equation}
 (\mathcal{D}^2 \phi)\Big|_{\mathcal{M}\times\{0,1\}}= D^2
\phi \Big|_{\mathcal{M}\times\{0,1\}}=(-\slashed{\nabla}^2+|m|^2)\phi \Big|_{\mathcal{M}\times\{0,1\}} .
\end{equation}

%%%%%%%%%%%%%%%%%%%%%%%%%%%%%%%%%%%
\subsection{Dirac operator and dispersion relation}
Let us investigate the relation between distance for a two-sheet
space and Dirac operator. To proceed, one needs to define the
notion of parallel transport for such a manifold.

\begin{Def} Let $\mathcal{M}\times\{0,1\}$ be a two-sheet space. A spinor field
    $\psi \in L^2(\mathcal{M})\otimes\mathbbm{C}^2$ is parallel transporting between $\mathcal{M}_i$ and $\mathcal{M}_j$
    (which form the two-sheet spacetime), if there exists a spinor field
    $\phi \in L^2(\mathcal{M}_5, S)$, such that $\phi(y,j)$ is the
    parallel transport of $\phi(x,i)$, for $(x,i), (y,j)\in
  \mathcal{M}_5$, and
\begin{equation}
(\mathcal{D}^2 \phi)\Big|_{\mathcal{M}\times\{0,1\}}= D^2
\phi \Big|_{\mathcal{M}\times\{0,1\}}=D^2\psi .
\label{paral-trans}
\end{equation}
\end{Def} 
\noindent
Note that, if the spinor $\phi$ exists, then its uniqueness is
guaranteed by the uniqueness of the solution of the differential
equation (geodesic equation in this case).

\begin{Def} A parallel transporting spinor field $\psi \in L^2(\mathcal{M})\otimes\mathbbm{C}^2$, with
  $(\psi,\psi)\not=0$, is called \textbf{causal} if
\begin{equation}
\frac{( D\psi, D \psi)}{(\psi,\psi)} \geq 0,
\end{equation}
and is \textbf{harmonic} if the equality holds. Otherwise, the spinor
is \textbf{non-causal}.
\end{Def}

In the following, we will relate the definition for a causal spinor to
the causal structure, Eq.~(\ref{causal-str}), in the case of an
almost-commutative geometry. 

\begin{Prop} Let $\psi \in
  L^2(\mathcal{M})\otimes\mathbbm{C}^2,~(\psi,\psi)\not=0$ be a parallel transporting spinor field between $ \mathcal{M}_i$ and $ \mathcal{M}_j$. The
  geodesic of the spinor connecting
  any two points $(x,i)$ and $(y,j)$ is null iff the spinor field is 
harmonic.\\\\
  \textbf{Proof} To prove this proposition, one has in principle to
  consider different cases. In the following, we will draw the proof
  for $i=0, j=1$. The other cases can be shown trivially.

  First suppose $\psi$ is a parallel transporting spinor field between
  $ \mathcal{M}_0$ and $ \mathcal{M}_1$. For any $(x,0), (y,1)\in
  \mathcal{M}_5$ there is a spinor $\phi \in L^2(M_5, S)$  such that $\phi(y,j)$ is the
    parallel transport of $\phi(x,i)$. 
    \\

a)  If the geodesic for $\phi(t,\mathbf{x},x_F)$ is null,
  then its line element is also null i.e.
\begin{equation}
dt^2 = |d\mathbf{x}|^2+\frac{1}{|m|^2}dx^2_F~.
\end{equation}
Since $dt^2$ and
$|d\mathbf{x}|^2+\frac{1}{|m|^2}dx^2_F$ are infinitesimal in
Euclidean space, one can write
\begin{align}
  \frac{\partial^2 \phi}{\partial t^2} =
  &\left(\sum^3_{i=1}\frac{\partial^2}{\partial
      x_i^2}+|m|^2\frac{\partial^2}{\partial x_F^2}\right)\phi
  ~. \label{eq:L-Dirac}
\end{align}
The restriction of Eq.~(\ref{eq:L-Dirac}) onto the two-sheet space
reads
\begin{align}
  \frac{\partial^2 \psi}{\partial t^2}=\frac{\partial^2 \phi}{\partial t^2} \Big|_{\mathcal{M}\times\{0,1\}}=&\left(\sum^3_{i=1}\frac{\partial^2}{\partial x_i^2}+|m|^2\frac{\partial^2}{\partial x_F^2}\right)\phi \Big|_{\mathcal{M}\times\{0,1\}}\nonumber\\
  =& \left(\sum^3_{i=1}\frac{\partial^2}{\partial
      x_i^2}+|m|^2\right)\psi~,
\end{align}
using Eq.~(\ref{paral-trans}).  Therefore,
\begin{align}
(D\psi,D\psi)=(\psi,D^+D\psi)=-
\Big(\phi|_{\mathcal{M}\times\{0,1\}},\mathcal{D}^2\phi|_{\mathcal{M}\times\{0,1\}}\Big)=
-\Big(\phi|_{\mathcal{M}\times\{0,1\}},\{-\slashed{\nabla}^2+D^2_F\}\phi
|_{\mathcal{M}\times\{0,1\}}\Big)=0~,
\end{align}
where we have used that that Dirac operator is Krein anti-self-adjoint.
\\
b) Conversely, assuming that the spinor on the two-sheet space is
harmonic,
\begin{equation}
  0=(D\psi,D\psi)=\Big({\cal D}\phi|_{\mathcal{M}\times\{0,1\}},{\cal D}\phi|_{\mathcal{M}\times\{0,1\}}\Big)
  =-\Big(\phi|_{\mathcal{M}\times\{0,1\}},\Big\{-\slashed{\nabla}^2
  -|m|^2\frac{\partial^2}{\partial x_F^2}\Big\}\phi|_{\mathcal{M}\times\{0,1\}}\Big)~. \label{eq:0-norm}
\end{equation} 
 Consider an inner product $(~,~)_5$ on $L^2(M_5,S)$ as
\begin{equation}
  ({\cal D}\phi,{\cal D}\phi)_5=-\Big(\phi,\Big\{-\slashed{\nabla}^2-|m|^2\frac{\partial^2}{\partial x^2_F}
  \Big\}\phi\Big)_5
  =-\int^0_1{dx_F}\Big(\phi(x_F),\Big\{-\slashed{\nabla}^2-|m|^2\frac{\partial^2}{\partial x^2_F}\Big\}
  \phi(x_F)\Big)~\label{eq:5-norm}.
\end{equation}
Then, using Eq.~(\ref{eq:0-norm}) and the fact that norm of a spinor
is preserved along a geodesic, the inner product (\ref{eq:5-norm})
vanishes, implying
\begin{align}
  \frac{\partial^2 \phi (x)}{\partial
    t^2}=&\left(\sum^3_{i=1}\frac{\partial^2}{\partial
      x_i^2}+|m|^2\frac{\partial^2}{\partial
      x_F^2}\right)\phi(x) ~,
\end{align}
at every point on the geodesic. The inverse of $\frac{\partial^2
}{\partial t^2}$ and of
$\sum^3_{i=1}\frac{\partial^2}{\partial
  x_i^2}+|m|^2\frac{\partial^2}{\partial x_F^2}$ give a line
element, which is null, therefore, the geodesic is itself null.
\end{Prop}

Let us note that in this study we restrict ourselves to the case of
harmomic spinors, the reason being that we want to investigate
their implications for the dispersion relation.  The next proposition
will show that harmonic spinors yield the energy-momentum
dispersion relation, meaning that they can be interpreted as physical
matter fields.

\begin{Prop} 
  Let $X$ be a compact subset of $\mathcal{M}$, and let $(A,\tilde{A},\mathcal{H},D)$ be the product of the Lorentzian
  spectral triple
  $(C^\infty(X),L^2(X,S),-i\slashed{\partial})$
  and the finite spectral triple $(A_F,\mathcal{H}_F,D_F)$. The
  eigenspinors $\Psi_n$ of the Dirac operator, with $(\Psi_n,\Psi_n)\not=0$, are harmonic iff
  their eigenvalues satisfy the energy-momentum
  dispersion relation.\\
  \textbf{Proof}

  Let $\Psi_n:=\psi_p\otimes e_i \in {\rm Dom}D$ be a normalised
  eigenspinor of $D$, where $\psi_p$ and $e_i$ are eigenstates of $
  \slashed{\partial}^2$ and $D^2_F$, respectively. 
  Note that, we choose the compact set $X\subset \mathcal{M}$ so that $\psi_p=\xi_p e^{i(-Et+\mathbf{p\cdot x})}$, 
for $\xi_p$ a constant spinor, is square integrable.
  We will distinguish
  two cases, namely whether $D^2_Fe_i=0$ vanishes or not.\\
  a) $D^2_Fe_i=0$\\
\begin{align}
  ( D\Psi_n, D \Psi_n)= (  \psi_p\otimes e_i,D^+ D  \psi_p\otimes e_i)=&~( \psi_p , \slashed{\partial}^2 \psi_p)(e_i,e_i)=(E^2-\mathbf{p}^2)( \psi_p , \psi_p ) \nonumber \\
  \Rightarrow \frac{( D\Psi_n, D \Psi_n)}{(\Psi_n,\Psi_n)} =&~E^2-\mathbf{p}^2~, \label{eq:ML}
\end{align}
where $-E^2$ denotes the eigenvalue of the $\partial^2/\partial t^2$
operator, and $-p_i^2$ stands for the eigenvalue of
$\partial^2/\partial x_i^2$. (${\bf p}$ denotes a three-vector.)

The r.h.s. of Eq.~(\ref{eq:ML}) is the energy-momentum dispersion
relation for a massless fermion iff $( D\Psi_n, D \Psi_n)=0$
i.e. $\Psi_n$ is harmonic.
\\
b) $D^2_Fe_i\not= 0$\\
\begin{align}
  ( D\Psi_n, D \Psi_n)=( \psi_p\otimes e_i,  D^+D \psi_p\otimes e_i)  =& ~(E^2-\mathbf{p}^2 )(\psi_p , \psi_p )( e_i , e_i) - m^2_i   (\psi_p , \psi_p )( e_i , e_i) \nonumber\\
  \Rightarrow \frac{(D\Psi_n, D \Psi_n)}{(\Psi_n,\Psi_n)}  =&
  ~E^2-\mathbf{p}^2-m^2_i~. \label{eq:EPdis}
\end{align}
Correspondingly, the r.h.s. of Eq.~(\ref{eq:EPdis}) is the
energy-momentum dispersion relation for a massive fermion iff $\Psi_n$
is harmonic.
\end{Prop}
 
Combining Propositions 2, 3 and 4 with Eq.~(\ref{def}), one may argue
that the energy-momentum dispersion relation has its origin in the
geometric construction of the almost-commutative manifold. Due to the
causal relation between the two sheets, one may interpret this
statement as the interaction between a fermion on one sheet and an
anti-fermion on the other one.

To highlight the validity of Proposition~4 in the case of inner
fluctuations of the Dirac operator, we will consider below a simple
toy model, namely electroweak theory with massless neutrinos.

%%%%%%%%%%%%%%%%%%%%%%%%%%%%%%%%%%%%%%%%%%%%%%%%%%%%%%%%%%%%%%%%%
\subsection{A toy model: Electroweak theory with massless neutrinos}

Consider the electroweak theory and assume neutrinos to be
massless. To explain this theory in the context of almost-commutative
spectral geometry, let us take the product of a Lorentzian spectral
triple $(C^\infty_0(\mathcal{M}),
L^2(\mathcal{M},S),-i\slashed{\partial})$ with a finite spectral triple
for the electroweak theory~\cite{Walter}. The spectral triple for the
discrete (internal) space $F$ is given by the algebra $A_F$, the
Hilbert space ${\cal H}_F$ and the Dirac operator $D_F$:
\begin{eqnarray}
  A_F&=&\mathbbm{C}\oplus\mathbbm{H}~,\\
  \mathcal{H}_F&=&\mathcal{H}_l\oplus\mathcal{H}_{\bar{l}}~,\\ 
  D_F&=&\left(\begin{array}{cccc}  0 &  Y^* & 0 & 0 \\
      Y &  0    &  0 & 0 \\
      0 &  0    &  0 & \bar{Y}^*\\
      0 &  0    &  \bar{Y} & 0 \end{array}\right)~, \label{eq:Dis-rac}
\end{eqnarray}
where $Y$ is a $2\times2$ mass matrix
\begin{equation}
Y=\left(\begin{array}{cc}  0 & 0 \\
                                             0 & m_e   \end{array}\right)~,
\end{equation}
with $m_e$ a complex parameter. 

Assuming all inner fluctuations to vanish, apart those of the scalar
field $\Phi$, the fluctuated Dirac operator for the almost-commutative manifold is
\begin{equation}
D_\Phi=-i\slashed{\partial}\otimes\mathbbm{I}_F+\gamma^5\otimes \Phi~,
\end{equation}
with
\begin{align}
\Phi=&D_F+a[D_F,b]+J_Fa[D_F,b]J^*_F \nonumber\\
  =& \left(\begin{array}{cc}  \phi & 0 \\
                                           0 & \bar{\phi}   \end{array}\right)~,
\end{align}
for $a,b\in C^\infty_0(\mathcal{M},A_F)$ and 
\begin{equation}
  \phi=\left(\begin{array}{cccc}  0 &   0 & 0 & 0 \\
      0 &  0    &  -\bar{m}_e h_2 & \bar{m}_e(h_1+1) \\
      0 &  -m_e\bar{h}_2    &  0 & 0\\
      0 &  m_e(\bar{h}_1+1)    &  0 & 0 \end{array}\right)~,
\end{equation}
where $h_1, h_2$ are complex functions. The trace of $\Phi^2$ is given by
\begin{equation}
{\rm Tr}\Phi^2=2|m_e|^2|\varphi|^2,
\end{equation}
where $\varphi:=(h_1+1, h_2)$ is a doublet. Assuming $\varphi$ undergoes symmetry breaking and denoting by $v$ the new VEV, we can choose $\varphi=(v+h,0)$, where $h$ is a small fluctuation around the vacuum.

To derive the dispersion relation, we will need $D_\Phi^2$, given by
\begin{align}
  D^2_\Phi=&~-\slashed{\partial}^2\otimes\mathbbm{I}_F+\gamma^\mu\gamma^5\otimes\partial_\mu \Phi
+\gamma^5\gamma^\mu \otimes\partial_\mu \Phi + \mathbbm{I}_4\otimes \Phi^2 \nonumber\\
  =&~-\slashed{\partial}^2\otimes\mathbbm{I}_F+ \mathbbm{I}_4\otimes \Phi^2 ~,
\end{align}
where we have used $\{\gamma^5, \gamma^\mu\}=0$.  
We denote the basis of ${\cal H}_l$ and ${\cal H}_{\bar l}$ by $\{\nu_R, e_R, \nu_L, e_L\}$ 
and $\{\bar{\nu}_R, {\bar e}_R, {\bar \nu}_L, {\bar e}_L\}$, respectively.

 The dispersion relation associated with harmonic
eigenspinors $\psi_p\otimes e_L$ and $\psi_p\otimes \nu_L$ (the same
result can be obtain for right-handed particles and anti-particles)
can be found as follows:
\begin{align}
(  \psi_p\otimes e_L,D^2_\Phi \psi_p\otimes e_L)=0~.
\end{align}
However,
\begin{align}
(  \psi_p\otimes e_L,D^2_\Phi \psi_p\otimes e_L )=&( \psi_p ,-\slashed{\partial}^2 \psi_p )(e_L,e_L)+( \psi_p , \psi_p)(e_L,\Phi^2 e_L) \nonumber\\
=&(-E^2+\mathbf{p}^2)( \psi_p , \psi_p )(e_L,e_L)+\|m_e\|^2(v^2+2vh+h^2)( \psi_p , \psi_p)(e_L, e_L) \nonumber\\
=&-E^2+p^2+\|m_e\|^2(v^2+2vh+h^2)~.
\end{align}
Hence,
\begin{equation}
E^2=p^2+\|m_e\|^2(v^2+2vh+h^2)~.
\end{equation}
Since the fluctuation is small, we have $E^2\sim p^2+\|m_e\|^2v^2$, which corresponds to the case b) in the proof of proposition~4. 
Similarly, the harmonic spinor $\psi_p\otimes \nu_L$  yields
\begin{equation}
E^2=p^2~,
\end{equation}
 corresponding to the case a) of the proof in proposition~4.

\section{Conclusions}
In the context of almost-commutative spectral geometry,
spectral distance between a pair of pure states in $M\times F$
was shown to be related to the
infinitesimal distance $ds^2$ between two points in $M$ and the
distance between internal states in $F$, via the Pythagorean
theorem~\cite{ncgdis}.  Such a relation was
shown~\cite{Martinetti:2009tr} also to be valid for $1/ds^2$. For the
latter case, one may observe a similarity between the Pythagorean
theorem and the energy-momentum dispersion relation, implying a
geometric origin of the dispersion relation. 

To confirm the above observation, one has to reformulate the inverse
distance, given by the inverse of the Dirac operator, in the context
of Lorentzian almost-commutative spectral geometry.  Following
Ref.~\cite{NCG-cone}, one can write down the spectral triple for a
Lorentzian almost-commutative manifold, and get the corresponding
Dirac operator.

Having the Lorentzian Dirac operator we are able to calculate the
 distance for a two-sheet manifold and define the notion of a
causal structure for such a geometry.
We were then able to show that the causal
structure on a flat almost-commutative space can be identified with
the causal structure on the five-dimensional Minkowski space with
metric
\begin{equation}
  g_{ab}=\left(\begin{array}{cc}  \eta_{\mu\nu} & 0 \\ 0 &
 1/|m|^2 \end{array}\right)~. \nonumber
 \end{equation}
 We have then suggested that spinors may be classified into causal, harmonic
 and non-causal ones.  The condition satisfied by harmonic spinors
 propagating in an almost-commutative manifold is equivalent to the
 causal relation, as suggested in Ref.~\cite{NCG-cone}.  We have further
 shown that a spinor is harmonic if and only if it satisfies the
 energy-momentum dispersion relation.

 We have hence shown the geometric origin of the dispersion relation
 in the context of almost-commutative spectral geometry.

 \acknowledgements This work was supported in part by the Action
 MP1405 QSPACE, from the European Cooperation in Science and
 Technology (COST). We thank W.~van suijlekom for organising the 
 conference `Gauge Theory and Noncommutative Geometry,' where 
 we had an opportunity to discuss and exchange interesting ideas. 
 A.~W. thanks M.~Eckstein for very helpful comments.

%\newpage

\end{document}